\documentclass[english,prd,twocolumn,superscriptaddress,nofootinbib,preprintnumbers]{revtex4}
\usepackage[latin1]{inputenc}
\usepackage{graphicx}
\usepackage{amssymb}

\makeatletter

\usepackage{amsfonts}
\usepackage{dcolumn}
\usepackage{hyperref}

\def\be{\begin{equation}}
\def\ee{\end{equation}}
\def\ba{\begin{eqnarray}}
\def\ea{\end{eqnarray}}
\def\bs{\begin{subequations}}
\def\es{\end{subequations}}

\newcommand{\tr}{\tilde{r}}

\usepackage{color}

\makeatother

\usepackage{babel}
\begin{document}

\title{Solar system and equivalence principle constraints \\
on $f(R)$ gravity by chameleon approach}

\author{Salvatore Capozziello}
\affiliation{Dipartimento di
Scienze Fisiche and INFN, Sez. di Napoli, Universita di Napoli
Federico II, Compl. Univ. di Monte S. Angelo, Edificio G, Via
Cinthia, I-80126 - Napoli, Italy}

\author{Shinji Tsujikawa}
\affiliation{Department of
Physics, Gunma National College of Technology, Gunma 371-8530,
Japan}

\begin{abstract}

We study constraints on $f(R)$ dark energy models from  solar
system experiments combined with experiments on  the violation of
equivalence principle. When the mass of an equivalent scalar
field degree of freedom is heavy in a region with high
density, a spherically symmetric body has a thin-shell
so that an effective coupling of the fifth force is suppressed
through a chameleon
mechanism. We place experimental bounds on the cosmologically
viable models recently proposed in literature which have an
asymptotic form $f(R)=R-\lambda R_c [1-(R_c/R)^{2n}]$ in the
regime $R \gg R_c$. From the solar-system constraints on the
post-Newtonian parameter $\gamma$, we derive the bound $n>0.5$,
whereas the constraints from the violations of weak and strong equivalence
principles give the bound $n>0.9$. This allows a possibility to
find the deviation from the $\Lambda$CDM cosmological model.
For the model $f(R)=R-\lambda R_c(R/R_c)^p$ with $0<p<1$
the severest constraint
is found to be $p<10^{-10}$, which shows that this model is hardly
distinguishable from the $\Lambda$CDM cosmology.

\end{abstract}

\date{\today}

\maketitle

The recent data coming from the luminosity distance of 
Supernovae Ia \cite{sneIa}, the wide galaxy surveys
\cite{lss} and the anisotropy of Cosmic Microwave Background
\cite{cmbr} suggest that about 70\,\% of the energy 
density of the present universe is composed by dark energy 
responsible for an accelerated expansion.
The cosmological constant is the most relevant candidate 
to interpret the cosmic expansion, but, in order to 
overcome its intrinsic shortcomings associated with the 
energy scale, several alternative models such as quintessence 
and k-essence have been proposed 
(see Ref.~\cite{review} for reviews).
Most of these models have the common feature to
introduce  new sources into the cosmological dynamics, but, from
an ``economic" point of view, it would be preferable to develop
scenarios consistent with observations without invoking extra
parameters or components non-testable  at a
fundamental level.

The simplest extension to the $\Lambda$CDM model is
presumably the so called $f(R)$ gravity,
where $f(R)$ is a general function of the Ricci 
scalar $R$ \cite{Capo} (see Ref.~\cite{Star80}
for an early work).
In Ref.~\cite{AGPT} the authors derived the conditions under
which a successful sequence of radiation, matter and accelerated
epochs can be realized. In addition the stability conditions
$f_{,R}>0$ and $f_{,RR}>0$ are required to
avoid ghosts and tachyons  for $R \ge R_1$, where $R_1$
is the Ricci scalar at a de-Sitter point \cite{Star}.
There exist viable $f(R)$ models that can satisfy both background
cosmological constraints
and stability conditions \cite{Li,AT07,Hu,Star,Appleby,Tsuji,Navarro,Odintsov,TUT}.
These models can satisfy solar system constraints 
under a chameleon mechanism, that is, 
a nonlinear effect arising from a large departure from the background 
value of $R$ \cite{Hu,Navarro,Tsuji,TUT}.
In this brief report, we place constraints on
viable $f(R)$ gravity models under the chameleon 
mechanism \cite{chame}
by using both solar-system and equivalence principle 
bounds.

We start with the following action in $f(R)$ gravity:
\begin{equation}
\label{action}
S=\int{\rm d}^{4}x\sqrt{-g}\,
f(R)/2+S_{m}(g_{\mu \nu}, \Psi_m)\,,
\end{equation}
where $S_m$ is a matter Lagrangian
that depends on the metric $g_{\mu \nu}$ and matter
fields $\Psi_m$.
We use the unit $M_{\rm pl}^2=(8\pi G)^{-1}=1$, where
$M_{\rm pl}$ and $G$ are a reduced Planck mass and
a bare gravitational constant respectively.

We introduce a new metric variable $\tilde{g}_{\mu \nu}$ 
and a scalar field $\phi$, as
\begin{eqnarray}
\tilde{g}_{\mu \nu}=\psi g_{\mu \nu}\,, \quad
\phi=\sqrt{3/2}\,{\rm ln}\,\psi\,,
\end{eqnarray}
where $\psi=\partial f/\partial R$.
Then the action in the Einstein frame is given
by \cite{Maeda}
\begin{eqnarray}
\label{actione}
S &=&\int{\rm d}^{4}x\sqrt{-\tilde{g}}\left[
\tilde{R}/2-(\tilde{\nabla}\phi)^2/2
-V(\phi) \right] \nonumber \\
& &+S_m (\tilde{g}_{\mu \nu} e^{2\beta \phi}, \Psi_m)\,,
\end{eqnarray}
where 
\begin{eqnarray}
\beta=-\frac{1}{\sqrt{6}}\,,\quad
V=\frac{R(\psi)\psi-f}{2\psi^2}\,.
\end{eqnarray}
The field $\phi$ is directly coupled to a non-relativistic 
matter with a constant coupling $\beta$.

In a spherically symmetric spacetime, 
the variation of the action (\ref{actione}) with 
respect to the scalar field $\phi$ gives
\begin{eqnarray}
\label{dreq}
\frac{{\rm d}^2 \phi}{{\rm d} \tr^2}+
\frac{2}{\tr} \frac{{\rm d}\phi}{{\rm d}\tr}=
\frac{{\rm d}V_{\rm eff}}{{\rm d}\phi}\,,
\end{eqnarray}
where $\tr$ is the distance from the center 
of symmetry and
\begin{eqnarray}
\label{Veff}
V_{\rm eff}(\phi)=V(\phi)+
e^{\beta \phi}\rho^*\,.
\end{eqnarray}
Here $\rho^*$ is a conserved quantity in the Einstein frame \cite{chame},
which is related with the energy density $\rho$
in the Jordan frame via the relation $\rho^*=e^{3\beta \phi}\rho$.

We assume that a spherically symmetric body has a constant
density $\rho^*=\rho_A^*$ inside the body ($\tr<\tr_c$)
and that the energy density outside the body ($\tr>\tr_c$)
is $\rho^*=\rho_B^*$. 
The mass $M_c$ of the body and the gravitational potential $\Phi_c$
at the radius $\tr_c$ are given by $M_c=(4\pi/3)\tr_c^3 \rho_A^*$
and $\Phi_c=M_c/8\pi \tr_c$, respectively.
The effective potential $V_{\rm eff} (\phi)$ has two minima 
at the field values $\phi_A$ and $\phi_B$
satisfying $V_{\rm eff}' (\phi_A)=0$ and 
$V_{\rm eff}' (\phi_B)=0$, respectively. 
The former corresponds to the region with a high density that gives rise
to a heavy mass squared $m_A^2 \equiv V_{{\rm eff}}''(\phi_A)$,
whereas the latter to the lower density region with a lighter mass
squared $m_B^2 \equiv V_{{\rm eff}}''(\phi_B)$.

In the high-density regime with a heavy field mass, it is known 
that the spherically symmetric body has a thin-shell under the 
chameleon mechanism. 
When the thin-shell develops inside the body, the following 
thin-shell parameter is much smaller than the order 
of unity \cite{chame}:
\begin{eqnarray}
\label{delrc}
\frac{\Delta \tr_c}{\tr_c}
=\frac{\phi_B-\phi_A}{6\beta \Phi_c}\,.
\end{eqnarray}
Solving Eq.~(\ref{dreq}) with appropriate boundary conditions, 
the field profile outside the 
body ($\tr>\tr_c$) is given by \cite{chame}
\begin{eqnarray}
\label{phir1}
\phi(\tr) \simeq -\frac{\beta_{\rm eff}}{4\pi}
\frac{M_c e^{-m_B(\tr-\tr_c)}}{\tr}+\phi_B\,,
\end{eqnarray}
where the magnitude of the effective coupling, $\beta_{\rm eff}=
(3\beta) (\Delta \tr_c/\tr_c)$, is much smaller than unity 
when the thin-shell is formed.

Let us study concrete $f(R)$ models that can satisfy
local gravity constraints as well as cosmological and stability 
conditions. Hu and Sawicki \cite{Hu} proposed the following model
\begin{eqnarray}
\label{mo1}
f(R)=R-\lambda R_c \frac{(R/R_c)^{2n}}{(R/R_c)^{2n}+1}\,,
\end{eqnarray}
whereas Starobinsky \cite{Star} proposed another viable model
\begin{eqnarray}
\label{mo2}
f(R)=R-\lambda R_c \left[ 1-\left( 1+R^2/R_c^2
\right)^{-n} \right]\,.
\end{eqnarray}
In both models $n$, $\lambda$ and $R_c$ are positive constants.
Since $f(R=0)=0$, the cosmological constant disappears in a flat
spacetime. Other $f(R)$ models with similar features have
been discussed in Refs.~\cite{Appleby,Odintsov,Tsuji}.
In these models a de-Sitter point responsible for the late-time
acceleration exists at $R=R_1~(>0)$,
where $R_1$ is derived by solving the equation
$R_1f_{,R}(R_1)=2f(R_1)$ \cite{AGPT}.
Note that $R_c$ is not much different from the 
present cosmological density $\rho_c \simeq 10^{-29}$\,g/cm$^3$.

In the region $R \gg R_c$ both models (\ref{mo1})
and (\ref{mo2}) behave as
\begin{eqnarray}
\label{fR}
f(R) \simeq R-\lambda R_c \left[1-
\left( R_c/R  \right)^{2n} \right]\,.
\end{eqnarray}
Inside and outside the spherically symmetric body 
the effective potential (\ref{Veff}) has minima at
$\phi_A \simeq -\sqrt{6}n \lambda (R_c/\rho_A)^{2n+1}$ and 
$\phi_B \simeq -\sqrt{6}n \lambda (R_c/\rho_B)^{2n+1}$, 
respectively. Since $\rho_A \gg \rho_B \gg \rho_c$ one has 
$|\phi_A| \ll |\phi_B| \ll 1$ and $\tilde{r} \simeq r$,
provided that $n$ and $\lambda$ are not much different 
from the order of unity.
In the following we omit the tilde for the quantity $r$. 
{}From Eq.~(\ref{delrc}) the thin-shell parameter is 
approximately given by 
\begin{eqnarray}
\frac{\Delta r_c}{r_c} \simeq 
n \lambda \left( \frac{R_c}{\rho_B} \right)^{2n+1}
\frac{1}{\Phi_c}\,.
\end{eqnarray}

Let us first discuss post-Newtonian solar-system constraints on 
the model (\ref{fR}).
In the weak-field approximation the spherically symmetric
metric in the Jordan frame is 
\begin{eqnarray}
{\rm d}s^2=-[1-2{\cal A}(r)]{\rm d}t^2
+[1+2{\cal B}(r)]{\rm d}r^2+r^2 {\rm d}
\Omega^2\,,
\end{eqnarray}
where ${\cal A}(r)$ and ${\cal B}(r)$ are the functions
of $r$. It was shown in Ref.~\cite{Faul} that under 
the chameleon mechanism the post-Newton parameter, 
$\gamma={\cal B}(r)/{\cal A}(r)$, 
is approximately given by 
\begin{eqnarray}
\gamma \simeq \frac{1-\Delta r_c/r_c}
{1+\Delta r_c/r_c}\,,
\end{eqnarray}
provided that the condition $m_B r \ll 1$ holds
on solar-system scales.
The present tightest constraint on $\gamma$ is 
$|\gamma-1| <2.3 \times 10^{-5}$ \cite{Will}, which 
translates into 
\begin{eqnarray}
\label{solar}
\frac{\Delta r_c}{r_c}<1.15 \times 10^{-5}\,.
\end{eqnarray}

For the model (\ref{fR}) the de-Sitter point corresponds to
$\lambda=x_1^{2n+1}/(2(x_1^{2n}-n-1))$, where 
$x_1=R_1/R_c$.
Using this relation together with $\Phi_c \simeq 2.12 \times 10^{-6}$
for the Sun, the bound (\ref{solar}) leads to 
\begin{eqnarray}
\label{cons2}
\frac{n}{2(x_1^{2n}-n-1)}\left(
\frac{R_1}{\rho_B} \right)^{2n+1}<2.4 \times 10^{-11}\,.
\end{eqnarray}
{}For the stability of the de-Sitter point we require
that $m=Rf_{,RR}/f_{,R}<1$ at $R=R_1$ \cite{AGPT}, which gives
the condition $x_1^{2n}>2n^2+3n+1$.
Hence the term $n/2(x_1^{2n}-n-1)$ in Eq.~(\ref{cons2})
is smaller than 0.25 for $n>0$.
Assuming that $R_1$ and $\rho_B$ are of the orders of the present
cosmological density $10^{-29}$ g/cm$^3$ and the
baryonic/dark matter density $10^{-24}$ g/cm$^3$
in our galaxy, respectively, we obtain the constraint
\begin{eqnarray}
\label{cons3}
n>0.5\,.
\end{eqnarray}
Thus $n$ does not need to be much larger than unity.
Hu and Sawicki derived the Ricci scalar $R$
as a function of $r$ by considering the density profile
of the Sun.
While we have obtained the bound (\ref{cons3}) without taking into
account such modifications, this bound is
consistent with the one derived by Hu and
Sawicki (see Eq.~(67) in Ref.~\cite{Hu}).

Let us also study the models of the type \cite{Li,Faul,AT07}
\begin{eqnarray}
\label{frpower}
f(R)=R-\lambda R_c(R/R_c)^p\,, \quad
0<p<1\,,
\end{eqnarray}
where $\lambda$ and $R_c$ are positive constants.
We do not consider the models with negative $p$,
because they suffer from instability problems of perturbations
associated with negative $f_{,RR}$ \cite{Dolgov,per}
as well as the absence of the matter-dominated
epoch \cite{APT}.
In this case the field $\phi_B$ is given by
$\phi_B=-(\sqrt{6}/2)\lambda p
(R_c/\rho_B)^{1-p}$.
Since the de-Sitter point, $x_1=R_1/R_c$, satisfies the
relation $\lambda=x_1^{1-p}/(2-p)$,
the bound (\ref{solar}) translates into 
\begin{eqnarray}
\frac{p}{2-p} \left( \frac{R_1}{\rho_B} \right)^{1-p}
<4.9 \times 10^{-11}\,.
\end{eqnarray}
Taking $R_1=\rho_1=10^{-29}$ g/cm$^3$ and
$\rho_B=10^{-24}$ g/cm$^3$,
we obtain the constraint
\begin{eqnarray}
\label{pbound1}
p < 5 \times 10^{-6}\,.
\end{eqnarray}
Hence the deviation from the $\Lambda$CDM model
is very small. 

Let us next place experimental bounds from
a possible violation of the equivalence principle (EP).
In doing so we shall discuss the thin-shell condition 
around the Earth under the chameleon
mechanism \cite{chame}.
The Earth has a radius $r_{\oplus}=6 \times 10^{3}$ km
with a mean density $\rho_{\oplus} \simeq 5.5$ g/cm$^3$.
The atmosphere exists in the region
$r_{\oplus}<r<r_{{\rm atm}}$
with a homogeneous density $\rho_{\rm atm} \simeq 10^{-3}$ g/cm$^3$.
The region outside the atmosphere ($r>r_{{\rm atm}}$)
has a homogenous density $\rho_G \simeq 10^{-24}$ g/cm$^3$.
Defining the gravitational potentials as
$\Phi_{\oplus}=\rho_{\oplus}r_{\oplus}^2/6$
and $\Phi_{\rm atm}=\rho_{\rm atm}r_{\rm atm}^2/6$,
we have that $\Phi_{\oplus} \simeq
5.5 \times 10^3 \Phi_{\rm atm}$
because $\rho_{\oplus} \simeq 5.5 \times 10^3 \rho_{\rm atm}$
and $r_{\oplus} \simeq r_{\rm atm}$.
Recalling the relation $\Delta r_{\rm atm}/r_{\rm atm}=
(\phi_G-\phi_{\rm atm})/(6\beta \Phi_{\rm atm})$,
where $\phi_G$ and $\phi_{\rm atm}$
correspond to the field values at the local minima
of the effective potential (\ref{Veff}) in the regions
$r>r_{\rm atm}$ and $r_{\oplus}<r<r_{{\rm atm}}$
respectively, we find
$\Delta r_{\oplus}/r_{\oplus}
\equiv -(\phi_G-\phi_{\rm atm})
/\sqrt{6} \Phi_{\oplus} \simeq 2.0 \times 10^{-4}
(\Delta r_{\rm atm}/r_{\rm atm})$.

When the atmosphere has a thin-shell then the
thickness of the shell ($\Delta r_{\rm atm}$) is smaller
than that of the atmosphere: $r_s=10$-10$^2$\,km.
Taking the value $r_s=10^2$\,km and $r_{\rm atm}
=6.5 \times 10^3$\,km, we obtain
$\Delta r_{\rm atm}/r_{\rm atm}<1.6 \times 10^{-2}$.
Hence the condition for the atmosphere 
to have a thin-shell is estimated as
\begin{eqnarray}
\label{thincon}
\frac{\Delta r_{\oplus}}{r_{\oplus}} \lesssim 10^{-6}\,.
\end{eqnarray}

Let us discuss solar system tests of  EP that makes use of
the free-fall acceleration of the Moon and the Earth toward
the Sun. The constraint on the difference of
two accelerations is given by
\begin{eqnarray}
\label{etamoon}
\eta \equiv 2\frac{|a_{\rm Moon}-a_{\oplus}|}
{a_{\rm Moon}+a_{\oplus}}<10^{-13}\,.
\end{eqnarray}

The Sun and the Moon have the thin-shells like the
Earth \cite{chame}, in which case the field profiles
outside the spheres are given as in Eq.~(\ref{phir1})
with the replacement of corresponding quantities.
We note that the acceleration induced by a fifth force
with the field profile $\phi(r)$ and the effective coupling
$\beta_{\rm eff}$ is $a^{\rm fifth}=
|\beta_{\rm eff}\phi(r)|$.
Then the accelerations $a_{\oplus}$ and $a_{{\rm Moon}}$
are \cite{chame}
\begin{eqnarray}
\hspace*{-1.0em}a_{\oplus} &\simeq& \frac{GM_{\odot}}{r^2}
\left[ 1+3 \left(\frac{\Delta r_{\oplus}}
{r_{\oplus}}\right)^2 \frac{\Phi_{\oplus}}{\Phi_{\odot}}
\right]\,,\\
\hspace*{-1.0em}a_{\rm Moon} &\simeq&
\frac{GM_{\odot}}{r^2}
\left[ 1+3 \left(\frac{\Delta r_{\oplus}}
{r_{\oplus}}\right)^2 \frac{\Phi_{\oplus}^2}
{\Phi_{\odot}\Phi_{\rm Moon}}\right]\,,
\end{eqnarray}
where $\Phi_{\odot} \simeq 2.1 \times 10^{-6}$,
$\Phi_{\oplus} \simeq 7.0 \times 10^{-10}$ and
$\Phi_{\rm Moon} \simeq 3.1 \times 10^{-11}$
are the gravitational potentials of Sun,
Earth and Moon, respectively.
Hence the condition (\ref{etamoon}) translates into
\begin{eqnarray}
\label{boep}
\frac{\Delta r_{\oplus}}{r_{\oplus}}<2 \times 10^{-6}\,,
\end{eqnarray}
which gives the same order of the upper bound
as in the thin-shell condition (\ref{thincon}) 
for the atmosphere.
The constraint coming from the violation of strong
equivalence principle \cite{Will} provides a
bound $\Delta r_{\oplus}/r_{\oplus}<10^{-4}$ \cite{chame},
which is weaker than (\ref{boep}).

Let us derive constraints on the models (\ref{mo1})
and (\ref{mo2}) under the bound (\ref{boep}).
On using the relation $|\phi_G|=\sqrt{6}n\lambda
(R_c/\rho_G)^{2n+1} \gg |\phi_{\rm atm}|$, we obtain
\begin{eqnarray}
n\lambda \left( \frac{R_c}{\rho_G} \right)^{2n+1}
< 10^{-15}\,.
\end{eqnarray}
Taking the similar procedure we have taken to reach
Eq.~(\ref{cons3}) from Eq.~(\ref{cons2}), we find
the following constraint
\begin{eqnarray}
\label{bound2}
n>0.9\,.
\end{eqnarray}
This is stronger than the bound (\ref{cons3})
derived from post-Newtonian tests in the solar system.

In the model (\ref{frpower}) the bound (\ref{boep})
leads to $\lambda p \left( R_c/\rho_G \right)^{1-p}
<10^{-15}$, which gives the constraint
\begin{eqnarray}
\label{pbound2}
p<10^{-10}\,.
\end{eqnarray}
Thus the model is required to be very close to the $\Lambda$CDM
model to satisfy the condition (\ref{boep}).

Let us next discuss constraints from fifth force experiments
that are carried out in a vacuum \cite{Will}.
Modeling a vacuum chamber as a sphere with radius $r_{\rm vac}$,
the energy density is given by $\rho(r)=0$ for $r<r_{\rm vac}$
and  $\rho(r)=\rho_{\rm atm}$ for $r>r_{\rm vac}$.
Inside the chamber we consider two identical bodies
of uniform density $\rho_c$, radius $r_c$
and total mass $M_c$.
If these bodies have thin-shells, their field profiles
are given by
\begin{eqnarray}
\label{phir}
\phi(r)=-\frac{\beta_{\rm eff}}{4\pi}
\frac{M_ce^{-r/r_{\rm vac}}}{r}+\phi_{\rm vac}\,,
\end{eqnarray}
where $\phi_{\rm vac}$ is the field value when
the mass squared of the field balances with the
curvature $r_{\rm vac}^{-2}$ of the chamber.
In Eq.~(\ref{phir}) we used the fact that the interaction
range $m_B^{-1}$ outside the bodies is
of the order of $r_{\rm vac}$ \cite{chame}.
The laboratory experiment constrains the coupling to be
$2\beta_{\rm eff}^2<10^{-3}$ \cite{Will},
which translates into the condition
\begin{eqnarray}
\label{rccon}
\frac{\Delta r_c}{r_c}<1.7 \times 10^{-2}\,.
\end{eqnarray}
Thus it is crucial to have thin-shells to satisfy the experimental bound.

We have $\Delta r_c/r_c \simeq -\phi_{\rm vac}/\sqrt{6}\Phi_c$
under the condition that $|\phi_{\rm vac}|$ is much larger than
the field value $|\phi_A|$ inside the bodies.
A typical test body used in Hoskins {\it et al.}
has a mass $M_c \sim 40$ g and a radius
$r_c \sim 1$\,cm \cite{Will}.
Hence the bound (\ref{rccon}) translates into
\begin{eqnarray}
\label{bo}
|\phi_{\rm vac}|<10^{-28}\,.
\end{eqnarray}

For the models (\ref{mo1}) and (\ref{mo2}) we obtain
$\phi_{\rm vac}$ in the region $R \gg R_c$:
\begin{eqnarray}
\phi_{\rm vac}=-\sqrt{6}n\lambda
\left[ \frac{R_c r_{\rm vac}^2}
{6n(2n+1)\lambda} \right]^{\frac{2n+1}{2n+2}}\,.
\end{eqnarray}
Then the constraint (\ref{bo}) gives
\begin{eqnarray}
C \left( r_{\rm vac}/R_1^{-1/2}
\right)^{\frac{2n+1}{n+1}}<10^{-28}\,,
\end{eqnarray}
where $C\equiv  \sqrt{6}n\lambda
\left[ \frac{1}{6n(2n+1)\lambda}\frac{R_c}{R_1}
\right]^{\frac{2n+1}{2n+2}}$.
{}From the relation $\lambda=x_1^{2n+1}/(2(x_1^{2n}-n-1))$ 
we find that $C$ is not larger than
the order of 0.1.
Using $R_1^{-1/2} \sim H_0^{-1} \sim 10^{28}$ cm,
we get the following constraint:
\begin{eqnarray}
\label{nlast}
n>0\,.
\end{eqnarray}
This is much weaker than the bounds
(\ref{cons3}) and (\ref{bound2}).

For the model (\ref{frpower}) the field value $\phi_{\rm vac}$
is given by
\begin{eqnarray}
\phi_{\rm vac}=-\frac{\sqrt{6}}{2}\lambda p
\left[ \frac{R_c r_{\rm vac}^2}{3\lambda p (1-p)}
\right]^{\frac{1-p}{2-p}}\,.
\end{eqnarray}
Making use of the relation $\lambda=x_1^{1-p}/(2-p)$
at the de-Sitter point, the condition (\ref{bo}) gives the bound
\begin{eqnarray}
\label{plast}
p<1.5 \times 10^{-2}\,.
\end{eqnarray}
Again this is much weaker than the bounds (\ref{pbound1})
and (\ref{pbound2}).

In summary we have found that the models (\ref{mo1}) and (\ref{mo2}) are
consistent with the present local gravity experiments for $n>0.9$, whereas 
the model (\ref{frpower}) is hardly distinguishable from the
$\Lambda$CDM cosmology because of the constraint $p<10^{-10}$.
These bounds are stronger than those derived by post-Newtonian tests in 
the solar system and are the main results of our paper.
The models (\ref{mo1}) and (\ref{mo2}) allow the possibility
to show appreciable deviations from the $\Lambda$CDM
model cosmologically around the present
epoch \cite{Hu,Star,Tsuji,TUT}.
It will be certainly of interest to find some signatures of
modified gravity in future high-precision local gravity
experiments such as the STEP \cite{STEP} or GAIA \cite{gaia} satellites as
well as in cosmological observations \cite{cosmoobser} such as the galaxy power
spectrum, Cosmic Microwave Background and weak lensing.

\vspace{0.5cm}

We acknowledge useful discussions and comments on the topics
with Y.~Fujii, A.~Stabile and A.~Troisi.
S.\,T. is supported by JSPS (Grant No.\,30318802).


\end{document}